\documentclass[conference]{IEEEtran}
\IEEEoverridecommandlockouts
\usepackage{cite}
\usepackage{amsmath,amssymb,amsfonts}
\usepackage{algorithmic}
\usepackage{graphicx}
\usepackage{textcomp}
\usepackage{xcolor}
\usepackage{multirow}
\usepackage[super]{nth}
\def\BibTeX{{\rm B\kern-.05em{\sc i\kern-.025em b}\kern-.08em
    T\kern-.1667em\lower.7ex\hbox{E}\kern-.125emX}}
\begin{document}

\title{Classification of Distraction Levels Using Hybrid Deep Neural Networks From EEG Signals
\footnote{{\thanks{This research was supported by the Challengeable Future Defense Technology Research and Development Program (912911601) of Agency for Defense Development in 2020.}
}}
}

\author{\IEEEauthorblockN{Dae-Hyeok Lee}
\IEEEauthorblockA{\textit{Dept. Brain and Cognitive Engineering} \\
\textit{Korea University} \\
Seoul, Republic of Korea \\
lee\_dh@korea.ac.kr} \\

\and

\IEEEauthorblockN{Sung-Jin Kim}
\IEEEauthorblockA{\textit{Dept. Artificial Intelligence} \\
\textit{Korea University} \\
Seoul, Republic of Korea \\
s\_j\_kim@korea.ac.kr} \\

\and

\IEEEauthorblockN{Yeon-Woo Choi}
\IEEEauthorblockA{\textit{Dept. Artificial Intelligence} \\
\textit{Korea University} \\
Seoul, Republic of Korea \\
yw\_choi@korea.ac.kr}
}

\maketitle

\begin{abstract}
Non--invasive brain--computer interface technology has been developed for detecting human mental states with high performances. Detection of the pilots' mental states is particularly critical because their abnormal mental states could cause catastrophic accidents. In this study, we presented the feasibility of classifying distraction levels (namely, normal state, low distraction, and high distraction) by applying the deep learning method. To the best of our knowledge, this study is the first attempt to classify distraction levels under a flight environment. We proposed a model for classifying distraction levels. A total of ten pilots conducted the experiment in a simulated flight environment. The grand--average accuracy was 0.8437 ($\pm$0.0287) for classifying distraction levels across all subjects. Hence, we believe that it will contribute significantly to autonomous driving or flight based on artificial intelligence technology in the future.
\end{abstract}

\begin{small}
\end{small}

\section{INTRODUCTION}
Brain--computer interface (BCI) allows users to communicate between human and external devices by recognizing human's status and intention \cite{suk2014predicting, issa2020emotion, won2017motion, chen2021multiattention, kwak2019error, liu2021assessing, kwon2019subject, zhang2017hybrid}. In particular, the non--invasive BCI system has advantages in that it does not require a surgical operation and costs less than other BCI systems. Since electroencephalogram (EEG) can reflect the user's intention and current state, the BCI system using EEG has been developed \cite{abibullaev2021systematic, kim2022rethinking, lee2019connectivity, gupta2019utility, jeong2020decoding}. In addition, advanced BCI has been applied in various domains, including diagnosis of Alzheimer’s disease \cite{thung2018conversion} and controlling external devices such as a robotic arm \cite{jeong2020brain} and a speller \cite{lee2018high}.

One of the interesting issues in the BCI domain is the detection of human abnormal mental states with robust performances \cite{lee2020continuous, hobson2005sleep, wu2019pilots, lim2020unified, borghini2014measuring, do2019neural, sonnleitner2014eeg}. In addition, the technology of autonomous driving or flight is required when the driver's or pilot's mental state changes from a normal state to an abnormal state. Since the driver's or pilot's mental state is directly related to the safety of passengers, the technology to accurately detect their mental state has been developed. Controlling an aircraft is a challenging task, and it consumes a number of energy \cite{yen2009investigation}. Among various abnormal mental states, distraction occurs when pilots cannot concentrate on flight control for many different variables. In addition, the CAA of New Zealand reported that pilots' distraction was the major reason among the various reasons for flight accidents \cite{ccakir2016real}.

EEG signals are important signals in the case of detecting human's mental states because they reflect the human's mental states directly \cite{jeong2019classification}. A few research groups have studied detecting distraction using only EEG signals. Sonnleitner \textit{et al}. \cite{sonnleitner2014eeg} described the impact of an auditory secondary task on drivers' mental states during a primary driving task. Provoked reaction time to brake lights and EEG alpha spindles were analyzed to describe distracted drivers. Brake reaction times and alpha spindle rate were significantly higher in distraction. Wang \textit{et al}. \cite{wang2015eeg} assessed the differences in behavioral performance and EEG activity when participants performed a lane--keeping driving task and a mathematical problem--solving task. Their system achieved 84.6 $\pm$ 5.8 \% and 86.2 $\pm$ 5.4 \% classification accuracies in detecting the participants' attention on the math and driving tasks, respectively. Li \textit{et al}. \cite{li2021temporal} developed a novel deep learning--based distraction detection approach based on both temporal and spatial information. The results showed that their proposed approach achieved an overall binary (distraction vs. non--distraction) classification accuracy of 0.92.

The main contributions are as follows. \textit{i}) We acquired distraction--related EEG signals under a flight environment with pilots. We measured EEG signals corresponding to a normal state (NS), low distraction (LD), and high distraction (HD) according to the difference in tasks. \textit{ii}) We propose hybrid deep neural networks for classifying distraction levels. We obtained the highest classification performance using our proposed model compared to the conventional models. To the best of our knowledge, this is the first attempt to classify distraction levels based on the deep learning architecture robustly.

The rest of this paper is organized as follows. In Section II, we introduce our experimental design and the explanation of the proposed model. In Section III, we present the experimental results. Finally, we conclude this paper in Section IV.\\

\section{MATERIALS AND METHODS}
\subsection{Subjects}
Ten pilots (S1--S10, aged 25.6 ($\pm$0.52)) participated in our experiment. All of them had flight experience (over 100 hr.) in the Taean Flight Education Center. They had no history of psychiatric or neurological disorders. We informed the entire experimental protocols to pilots. Our experimental protocols and environment were approved by the Institutional Review Board of Korea University [1040548--KU--IRB--18--92--A--2]. Before the experiment, they consented according to the Declaration of Helsinki. For evaluating the experimental paradigm, we instructed the pilots to fill out a questionnaire to check their mental and physical conditions after finishing the experiment.\\

\subsection{Experimental Environment}
The Cessna 172 (Garmin, Olathe, KS) was used in the flight simulator which included the screen, the cockpit, a keypad, and the signal amplifier. A flight yoke and other control panels were the components of the cockpit for constructing a realistic flight environment. The subjects inputted the number using a keypad. We attached a keypad to the flight yoke directly for preventing the unnecessary movement when the subjects press the keypad. The signal amplifier (BrainAmp, Brain Products GmBH, Germany) was used to measure EEG and EOG signals. We set the sampling frequency of EEG and EOG signals as 1,000 Hz, and we used a 60 Hz notch filter for removing the power supply noise. We placed the 30 EEG channels according to the international 10/20 system. Also, we attached the four EOG channels to the horizontal and vertical lines around the eye. FCz and AFz channels were used to the reference and ground electrodes, respectively. We set up the impedance of electrodes below 10 k$\Omega$ by injecting the conductive gel.\\

\subsection{Experimental Protocol and Paradigm}
We designed the experimental paradigm for acquiring the distraction-based EEG signals effectively, as shown in Fig. 1. The tasks given to the subjects were not difficult for the pilot, as it is hard to obtain high--quality distraction--based EEG signals if the tasks are difficult. The air traffic control (ATC) message (length: 4--22 words) was presented to the subjects. The subjects simultaneously counted the number of words contained in the ATC message while performing the flight. We instructed the subject to count the number of words without any body movement. We divided the distraction level according to the length of the ATC message (level 1: 4--9 words, level 2: 10--14 words, level 3: 15--22 words). We defined level 1 as LD and level 2 and level 3 as HD.\\

\begin{figure}[t!]
\centering
\scriptsize
\centerline{\includegraphics[width=0.87\columnwidth, height=0.13\textwidth]{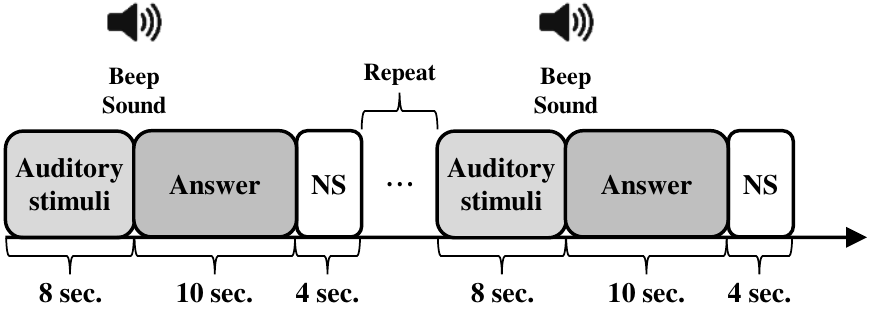}}
\caption{Experimental paradigms for inducing distraction in a simulated flight environment.}
\end{figure}

\subsection{Signal Preprocessing}
We conducted the preprocessing of EEG signals using a BBCI toolbox in MATLAB 2019a \cite{blankertz2010berlin}. We utilized the band—pass filter using a $2^{nd}$ order zero--phase Butterworth filter between 1 and 50 Hz and down—sampled the signals from 1,000 to 100 Hz. We applied the independent component analysis \cite{jung1998extended} for removing the contaminated components. Each trial was segmented into 1 sec. data without overlap \cite{jeong2019classification}. 1,860 samples ([Level 1] 10 samples$\times$40 trials + [Level 2] 10 samples$\times$40 trials + [Level 3] 10 samples$\times$40 trials + [Rest after entering the numbers of words] 4 samples$\times$120 trials + [Rest between $1^{st}$ and $2^{nd}$ sessions] 180 samples) were obtained for each subject. Across all subjects, 18,600 samples were obtained. Since there is a difference in the number of samples in the NS, LD, and HD, We solved the problem of data unbalance according to the number of samples in the LD with the smallest number of samples.\\

\subsection{Proposed Model}
We proposed the model to classify distraction levels using EEG signals accurately. EEG signals have various features such as spectral, spatial, and temporal information. Hence, we designed the model with the principle of a hybrid deep learning framework. The components of our proposed model were five convolutional blocks and one LSTM block.

The five convolutional blocks were utilized for extracting the significant spatial and spectral features from EEG signals. The $1^{st}$, $2^{nd}$, and $3^{rd}$ convolutional blocks have two layers, 1 $\times$ 5 filters, 1 $\times$ 1 stride, and a batch normalization layer. $4^{th}$ and $5^{th}$ convolutional blocks have three layers, 5 $\times$ 1 and 3 $\times$ 1 filters, respectively, 1 $\times$ 1 stride, and a batch normalization layer. In addition, we applied the maximum-- and average--pooling layers for avoiding the overfitting problem. Also, we used the exponential linear unit as an activation function. 

LSTM network, which is one of the recurrent neural networks, is an effective network for recognizing mental states. One LSTM block, which has two LSTM layers with 256 and 128 hidden units, respectively, was utilized for extracting the significant temporal features from EEG signals.

The last part of our proposed model is the classification block, which has three fully connected layers and a softmax layer. The hidden units of $1^{st}$ and $2^{nd}$ fully connected layers were 128 and 64, respectively. Also, the output of the $3^{rd}$ fully connected layer was fed to 3--way softmax.\\

\begin{table}[t!]
\centering
\caption{Comparison of the classification performances with the statistical analysis for classifying distraction levels among the conventional models and the proposed model.}
\tiny
\renewcommand{\arraystretch}{1.4}
\resizebox{\columnwidth}{!}{
\begin{tabular}{cccc}
\hline
Subject & PSD--SVM \cite{zhang2017design}        & DeepConvNet \cite{schirrmeister2017deep}    & Proposed        \\ \hline
S1      & 0.6831          & 0.7244          & 0.8128          \\
S2      & 0.6806          & 0.7276          & 0.8489          \\
S3      & 0.6184          & 0.7641          & 0.8403          \\
S4      & 0.6999          & 0.7575          & 0.8374          \\
S5      & 0.7305          & 0.7983          & 0.8401          \\
S6      & 0.7177          & 0.8012          & 0.9088          \\
S7      & 0.5804          & 0.7001          & 0.8235          \\
S8      & 0.7198          & 0.7798          & 0.8554          \\
S9      & 0.7087          & 0.7596          & 0.8612          \\
S10     & 0.6582          & 0.7341          & 0.8082          \\ \hline
Avg.    & 0.6797          & 0.7547          & \textbf{0.8437} \\
Std.    & 0.0483          & 0.0331          & \textbf{0.0287} \\ \hline
\textit{p}--value & \textless{}0.05 & \textless{}0.05 & -               \\ \hline
\end{tabular}}
\end{table}

\section{RESULTS AND DISCUSSION}
\subsection{Performance Evaluation}
We applied the five--fold cross--validation method for evaluating performance fairly. The dataset was randomly shuffled and divided into five parts. The four parts were used as the training set and the one part was used as the test set. Table I presented the comparison of the performances for classifying distraction levels among the conventional models and the proposed model. The conventional models used for performance comparison were the power spectral density--SVM (PSD--SVM) \cite{zhang2017design} and DeepConvNet \cite{schirrmeister2017deep}. The PSD--SVM is a model that uses PSD of the \textit{$\delta$}-- (1--4 Hz), \textit{$\theta$}-- (4--8 Hz), \textit{$\alpha$}-- (8--13 Hz), and \textit{$\beta$}--bands (13--30 Hz) as a feature and SVM as a classifier. The DeepConvNet is one of the representative models for decoding EEG signals. It consists of a total of four convolutional blocks. The first convolutional block consists of consecutive temporal and spatial convolutions, and the remaining convolutional blocks consist of temporal convolutions. Our model showed the highest average accuracy of 0.8437 ($\pm$0.0287) compared to the conventional models. In the case of our proposed model, S6 and S10 represented the highest and the lowest performance, respectively, and the values were 0.9088 and 0.8082, respectively. In addition, our proposed model indicated the lowest--standard deviation of 0.0287 compared to the conventional models. It means the highest stability across all subjects.

\begin{figure}[t!]
\centering
\scriptsize
\centerline{\includegraphics[width=\columnwidth, height=0.48\textwidth]{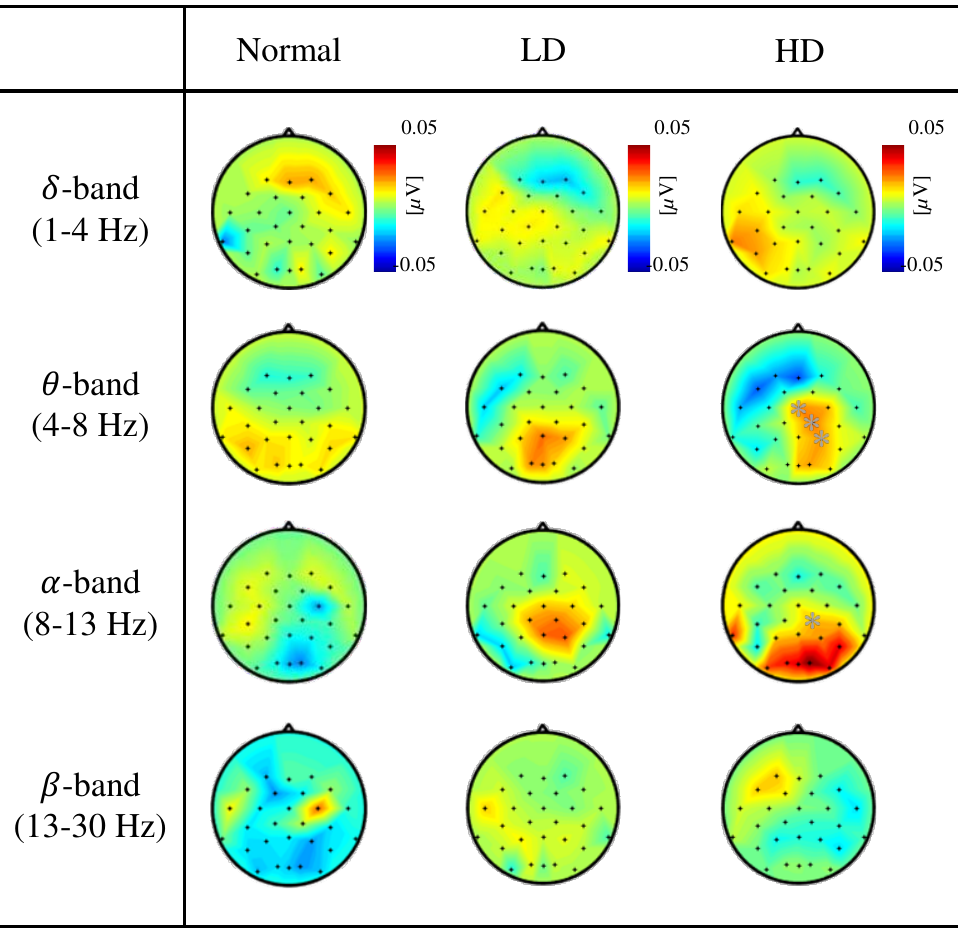}}
\caption{Scalp topographies according to the spectral bands (\textit{$\delta$}--, \textit{$\theta$}--, \textit{$\alpha$}--, and \textit{$\beta$}--bands) across the representative subject (S6). The locations of channels with the statistical significance are indicated as grey `$\ast$' (*: \textit{p}$<$0.05).}
\end{figure}

\subsection{Neurophysiological Analysis from EEG Signals}
We divided the brain region into temporal, central, and parietal regions and the frequency into \textit{$\delta$}--, \textit{$\theta$}--, \textit{$\alpha$}--, and \textit{$\beta$}--bands, as shown in Fig. 2. We represented the scalp topographies using EEG signals of the representative subject (S6). We used the grand--average band power to plot the scalp topographies. All EEG channels and each frequency band were used for calculating the amplitude. The amplitude was significantly different for each brain region and each spectral band. When distraction level increased, the amplitude of the \textit{$\theta$}--band in the centro--parietal region increased, and that of the \textit{$\alpha$}--band in the central region increased. The grey $\ast$ indicates the locations of channels with the statistical significance (\textit{p}$<$0.05). However, we could not find any particular spatial tendencies in the \textit{$\delta$}-- and \textit{$\beta$}--bands.

\section{CONCLUSION}
The pilot's abnormal mental states are the most important among the various reasons for flight accidents. Since it is directly related to the safety of passengers, detection of the pilots' mental state with robust performance is necessary. Detection of various mental states using EEG signals is one of the important challenging issues in the BCI domain. Especially, accurate classification of levels in--depth for mental states helps to prevent accidents caused by human error. In this paper, we proposed a model for classifying distraction levels using EEG signals. Our proposed model could classify distraction levels with robust performance compared to other conventional models. Since we acquired distraction--based EEG signals from highly disciplined pilots, we believe that our study will help advance autonomous driving or flight. To apply our study to the real--world environment, we will develop our proposed model after acquiring more various abnormal mental states-based EEG signals using our modified experimental paradigms.\\

\bibliographystyle{IEEEtran}
\bibliography{REFERENCE}

\end{document}